\documentstyle[12pt,epsfig]{article}

\begin{document}
\begin {center}
{\bf {\Large
Semihadronic and hadronic decays of $\rho^0$
and $\omega$ mesons coherently photoproduced in the isoscalar nuclei} }
\end {center}
%\smallskip
%\medskip
\begin {center}
Swapan Das \footnote {email: swapand@barc.gov.in} \\
{\it Nuclear Physics Division,
Bhabha Atomic Research Centre  \\
Mumbai-400085, India }
\end {center}
%\medskip

\begin {abstract}
The interference of the $\rho^0$ and $\omega$ mesons has been studied
in the hadronic and leptonic decay channels, i.e., dipion and
dilepton decay channels respectively, but this interference is not studied
yet in the semihadronic or semileptonic decay channel, e.g.,
$V \to \pi^0\gamma$. $V$ denotes either $\rho^0$ or $\omega$ meson.
To
look for the quoted interference in the $\pi^0\gamma$ decay channel as well
as the contribution of $\rho$ meson to the cross section,
the
correlated $\pi^0\gamma$ invariant mass distribution spectra are calculated
in the photonuclear reaction in the multi-GeV region.
It
is assumed that these bosons arise in the final state due to the decay of
$\rho^0$ and $\omega$ mesons which are photoproduced coherently in the
isoscalar nucleus. The elementary reaction in the nucleus is considered to
proceed as $\gamma N \to VN$; $V \to \pi^0\gamma$.
The
forward propagation of the $\rho$ and $\omega$ mesons and the near forward
emission of pion are considered, so that they can be described by the
eikonal form. The meson nucleus interactions are evaluated using $t\varrho$
approximation.
Replacing
the decay vertex $V \to \pi^0\gamma$ by $V \to \pi^+\pi^-$ in the above
formalism, it is also used to study the $\rho-\omega$ interference in the
$\pi^+\pi^-$ decay channel.
\end {abstract}

Keywords:
photonuclear reaction, $\rho-\omega$ interference,
meson nucleus potential

PACS number(s): 25.20.-x, 13.60.Le, 13.20.-v

\section{Introduction}

It is well known that the dipion (i.e., $\pi^+\pi^-$) and dilepton (e.g.,
$e^+e^-$, $\mu^+\mu^-$, ... etc.) emission in the reactions (multi-GeV
region) occur because of the decay of $\rho^0$ and $\omega$ mesons,
produced in the intermediate state of the reactions. The data of these
reactions are described better because of the interference of $\rho$
and $\omega$ mesons \cite{bauer}.
The
$e^+e^- \to \pi^+\pi^-$ reaction has been understood as
$e^+e^- \to \gamma \to V \to \pi^+\pi^-$ \cite{conn} (see the references
there in). The symbol $V$ is used to describe the vector meson, i.e.,
either $\rho^0$ or $\omega$ meson, throughout the text.
Though
the dominant contribution to the cross section of the
$e^+e^- \to \pi^+\pi^-$ reaction arises due to the $\rho$ meson, the data
are better reproduced because of the inclusion of $\omega$ meson
contribution to the reaction. The $\rho -\omega$ mixing parameter and pion
form factor \cite{conn} have been described well in the study of this
reaction.

The distinct $\rho-\omega$ interference has also been seen in the dilepton
production in the photonuclear reaction in the multi-GeV region
\cite{alven, biggs}:
$\gamma A \to VA$; $ V \to e^+e^- $. Various quantities, such as
$\gamma - V$ coupling constant, the relative phase between the $\rho$ and
$\omega$ mesons production, ... etc., have been extracted from the study of
this reaction.
The
$\rho-\omega$ interference was also studied in the $e^+e^-$ production (due
to the decay of the vector mesons produced near threshold) in the
photo-induced reaction on nucleon \cite{lutz}. Recently, the measurement on
the dilepton emission was reported from JLab to the search the quoted
interference in the $\gamma$ proton reaction \cite{djal}.
To
be mentioned, the $\rho^0$ meson photo-produced in the multi-GeV nuclear
reaction was detected by the $\pi^+\pi^-$ \cite{behr0} since the decay
branching ratio of $\rho^0 \to \pi^+\pi^-$ is $\sim 100\%$.
Though
the branching ratio of $\omega$ meson to the $\pi^+\pi^-$ decay channel is
very small compared to that of $\rho^0$ meson, the $\rho-\omega$
interference is also visible in the $\pi^+\pi^-$ photoproduction reaction on
nuclei \cite{biggs2, behr, alven2}.

It must be mentioned that the $\rho-\omega$ interference has been well
studied in the hadronic and leptonic decay channels but it is not yet done
for the semi-hadronic (or semi-leptonic) decay channels, i.e.,
$V \to \pi^0\gamma$ for example.
Recent
past, the correlated $\pi^0\gamma$ emission (in the GeV region) was studied
in the photonuclear reaction to investigate the modification of $\omega$
meson in the nucleus \cite{das}. However, the contribution of $\rho$ meson
to this reaction is never incorporated because (as shown latter) it is too
small in the $\omega$ meson peak region.

As stated earlier, the contribution of $\omega$ meson to the $\pi^+\pi^-$
production in the $e^+e^-$ and $\gamma A$ reactions is negligibly small
(compared to that of $\rho^0$ meson) but the data are better reproduced
due to the $\rho-\omega$ interference \cite{conn, behr}. In fact, this
would have never been known unless the contribution of $\omega$ meson 
included in the calculation.
Therefore,
the contribution of $\rho$ meson to the $\pi^0\gamma$ production reaction
should be studied so that the change in the cross section of the reaction
because of the $\rho$ meson (and $\rho-\omega$ interference) can be
recorded.
To
disentangle it, both $\rho^0$ and $\omega$ mesons are included in the
calculation for the cross section of the $\pi^0\gamma$ invariant mass
distribution in the photonuclear reaction,
i.e.,
the $\pi^0$ and $\gamma$ bosons in the final state are considered to arise
due to the decay of both $\rho^0$ and $\omega$ mesons produced coherently
in the intermediate state of the reaction.
The
elementary reaction occurring in the nucleus is assumed to proceed as
$\gamma N \to VN$; $V \to \pi^0\gamma$, where $V$ denotes the vector meson
(i.e., $\rho$ or $\omega$ meson, as mentioned earlier).

The forward production of the vector meson in the reaction is considered
since it ensures the coherence of its production amplitude in a nucleus
\cite{biggs, biggs2}. The emission of pion is also considered near forward
direction.
Both
the vector meson propagator and the pion distorted wave function are
expressed by the eikonal form. The meson nucleus interaction (optical
potential) is evaluated using $t\varrho$ approximation.
The
isoscalar nuclei are consider in the reaction so that one pion exchange
contribution to the vector meson (specifically, $\omega$ meson) production
in the nucleus can be ignored \cite{biggs, biggs2}.
Considering
the decay channel $V \to \pi^+\pi^-$ instead of $V \to \pi^0\gamma$ in the
reaction mechanism stated above, it can be used to describe the
$\pi^+\pi^-$ emission due to the decay of the vector meson produced
coherently in the photonuclear reaction.
Therefore,
the formalism is developed to calculate the cross section for both
$\pi^0\gamma$ and $\pi^+\pi^-$ invariant mass distribution spectrum in the
quoted reaction.

\section{Formalism}

The $(\gamma, V \to ab)$ reaction on a nucleus consists of three parts:
(i) the vector meson $V$ photoproduction in the nucleus,
(ii) the propagation of this meson through the nucleus, and
(iii) the decay of vector meson $V$ into $ab$ (i.e., $V\to ab$) in the
final state, where $ab$ represents either $\pi^0\gamma$ or $\pi^+\pi^-$.
The
first part can be described by the generalized potential or self-energy
of the vector meson in a nucleus \cite{das2}, i.e.,
\begin{equation}
\Pi_{\gamma A \to VA} ({\bf r}) =
K {\tilde f}_{\gamma N \to VN} (0) \varrho({\bf r}),
\label{pia}
\end{equation}
with $K = -4\pi E_V (1/{\tilde E}_V + 1/{\tilde E}_N)$.
${\tilde f}_{\gamma N \to VN}$ is the amplitude of the elementary reaction:
$\gamma N \to VN$. The symbol ``$tilde$'' on the quantities denote those
in the $\gamma N$ cm system. $\varrho ({\bf r})$ represents the matter
density distribution of the nucleus.
As
mentioned earlier, one pion exchange contribution can be neglected for
the vector meson production in the isoscalar nucleus \cite{biggs, biggs2}.

The propagator of the vector meson can be expressed as 
$ (-g^\mu_\nu + \frac{1}{m^2} k^\mu_V k_{V,\nu} )
G_V ( m; {\bf r-r^\prime} ) $ \cite{das, das3}, where the scalar part of it,
i.e., $ G_V ( m, {\bf r-r^\prime} ) $, describes the propagation of this
meson from its production point ${\bf r}$ to decay point ${\bf r^\prime}$.
For
the forward going vector meson, $G_V (m, {\bf r-r^\prime})$  can be well
described by the eikonal form, i.e.,
\begin{equation}
G_V (m; {\bf r - r^\prime})=
\delta ({\bf b-b^\prime}) \theta (z^\prime - z)
e^{i {\bf k}_V . {\bf (r-r^\prime) } }
D_{{\bf k}_V} (m; {\bf b}, z^\prime, z).
\label{proe}
\end{equation}
The factor $ D_{{\bf k}_V} (m; {\bf b}, z^\prime, z) $ in this equation
carries the information about the nuclear effect on the vector meson
during its propagation through the nucleus, since it involves the vector
meson nucleus optical potential $V_{OV}$. The expression for
$ D_{{\bf k}_V} (m; {\bf b}, z^\prime, z) $ is
\begin{equation}
D_{{\bf k}_V} (m; {\bf b}, z^\prime, z) = - \frac{i}{2k_{V\parallel}}
exp \left [ \frac{i}{2k_{V\parallel}} \int^{z^\prime}_z dz^{\prime \prime}
\{ \tilde {G}^{-1}_{0V}(m) -2 E_V V_{OV} ({\bf b}, z^{\prime \prime}) \}
\right ],
\label{prdf}
\end{equation}
where   $ \tilde {G}^{-1}_{0V}(m) = m^2-m^2_V+im_V\Gamma_V(m) $ is the
inverse of the free space vector meson propagator. $m_V$ is the pole mass of
this meson: $m_{\rho^0}=775.26$ MeV and $m_{\omega}=782.65$ MeV, listed in
Ref.~\cite{olive}. All other symbols carry their usual meanings.

The $\gamma$ wave function is described by the plane wave associated with
its polarization vector. As mentioned earlier, the particles $ab$ in the
final state (i.e., the decay products of the vector meson) are considered
either $\pi^+\gamma$ or $\pi^+\pi^-$.
For
the pion emission near the forward direction, the wave function for it can
be written in eikonal form \cite{das, glub}, i.e.,
\begin{equation}
\chi^{(-)*} ({\bf k}_\pi, {\bf r^\prime}) =
e^{ -i{\bf k}_\pi \cdot {\bf r^\prime} }
D^{(-)*}_{ {\bf k}_\pi}  ( {\bf b}, z^\prime ),
\label{pwfn}
\end{equation}
where $ D^{(-)*}_{ {\bf k}_\pi } ( {\bf b}, z^\prime ) $ denotes the
distortion due to the pion nucleus interaction (i.e., pion nucleus optical
potential $V_{O\pi}$). It is given by
\begin{equation}
D^{(-)*}_{ {\bf k}_\pi } ( {\bf b}, z^\prime )= exp
\left [ -\frac{i}{v_{\pi \parallel}}
\int^\infty_{z^\prime} dz_\jmath V_{O\pi} ({\bf b},z_\jmath) \right ],
\label{dspw}
\end{equation}
where $v_\pi$ is the velocity of pion.

The $T$-matrix $T_{fi}$ of the coherent ($\gamma, V \to ab$) reaction on a
nucleus is related to its reduced matrix ${\cal M}_{fi}$ as
$ T_{fi}=
\frac { {\cal M}_{fi} }
{ \sqrt { (2E_\gamma 2E_a 2E_b) } } $,
where ${\cal M}_{fi}$ can be written as
\begin{equation}
{\cal M}_{fi}=
\sum_{V=\rho^0,\omega} \Gamma_{Vab} F_{\gamma, V \to ab}.
\label{mxfi}
\end{equation}
$\Gamma_{Vab}$ in this equation describes $V \to ab$ decay vertex. The
Lagrangian for it is given latter.
$F_{\gamma, V\to ab}$ represents the space part of ${\cal M}_{fi}$, i.e.,
\begin{equation}
F_{\gamma, V\to ab}=
<ab| G_V (m; {\bf r - r^\prime}) \Pi_{\gamma A \to VA} ({\bf r}) |\gamma>.
\label{fve0}
\end{equation}
$G_V (m; {\bf r - r^\prime})$ and the wave function for the particles
$a$ and $b$ are given in Eqs.~(\ref{proe}) and Eq.~(\ref{pwfn})
respectively.

The cross section of the coherent $(\gamma, V\to ab)$ reaction on a nucleus
is given by
\begin{equation}
d\sigma = \frac{(2\pi)^4}{v_{\gamma A}} \delta^4 (k_i-k_f)
<|T_{fi}|^2> \Pi_{f(= a,b, A^\prime)} [d{\bf k}/(2\pi)^3]_f,
\label{dsc}
\end{equation}
where
$v_{\gamma A}$ is the relative velocity of the incident $\gamma$ with
respect to the target nucleus $A$, i.e., $ v_{\gamma A} = |v_\gamma-v_A |$.
The primed quantity, i.e., $A^\prime$, denotes the recoil nucleus. The
annular bracket around $T_{fi}$ represents the average over initial states
and the summation over final states.

\section{Result and Discussions}

The total decay width of the vector meson $\Gamma_V(m)$, appearing below
Eq.~(\ref{prdf}), is composed of the partial widths of its various decay
channels \cite{olive}:
$ \Gamma_{\rho^0} (m) =
99.94 \times 10^{-2} \Gamma_{\rho^0 \to \pi^+\pi^-} (m) +
6.0 \times 10^{-4} \Gamma_{\rho^0 \to \pi^0\gamma} (m) $,
and
$ \Gamma_\omega (m) =
89.9 \times 10^{-2} \Gamma_{\omega \to \pi^+\pi^-\pi^0} (m) +
8.5 \times 10^{-2} \Gamma_{\omega \to \pi^0\gamma} (m) +
1.6 \times 10^{-2} \Gamma_{\omega \to \pi^+\pi^-} (m) $.
The partial decay widths are illustrated in Ref. \cite{das4}.

The meson nucleus optical potentials $V_{OM}$ (i.e., $V_{OV}$ in
Eq.~(\ref{prdf}) and $V_{O\pi}$ in Eq.~(\ref{dspw})) are evaluated
using $t\varrho$ approximation \cite{glub, das6}. According to it,
$V_{OM}$ is given by
\begin{equation}
V_{OM} ({\bf r})
= -\frac{v_M}{2} [i+\alpha_{MN}] \sigma^{MN}_t \varrho ({\bf r}),
\label{opms}
\end{equation}
where $v_M$ is the velocity of the meson $M$. $\varrho ({\bf r})$ has been
defined in Eq.~(\ref{pia}).
The
scattering parameters $\alpha_{MN}$ and $\sigma^{MN}_t$ denote the ratio
of the real to imaginary part of the meson nucleon forward scattering
amplitude $f_{MN} (0)$ and the meson nucleon total cross section
respectively.

The imaginary part of $f_{MN}$ for the vector meson is extracted from the
data of the elementary $\gamma N \to VN$ reaction using vector meson
dominance (VDM) model \cite{kon, sibir, lyka}.
The
typical values of $\sigma^{VN}_t$ (in the considered energy region, i.e.,
3-5 GeV) at $k_V=4$ GeV/c are $\sigma^{\rho p}_t \simeq 35$ mb \cite{kon}
and $\sigma^{\omega p}_t \simeq 40$ mb \cite{lyka}.
The
real part of the $\rho^0$ meson nucleon scattering amplitude $f_{\rho N}$
is taken from the calculation of Kondratyuk et al., \cite{kon} which
reproduce the data in the quoted energy region.
For
the $\omega$ meson nucleon scattering, $\alpha_{\omega N}$ have been
calculated using additive quark model and Regge theory \cite{sibir}:
$ \alpha_{\omega N}
= \frac{0.173(s/s_0)^\epsilon - 2.726(s/s_0)^\eta}
       {1.359(s/s_0)^\epsilon + 3.164(s/s_0)^\eta} $,
with $s_0=1$ GeV$^2$, $\epsilon = 0.08$ and $\eta = -0.45$.
In
fact, the vector meson dominance (VMD) model relates $f_{MN}$ to
$f_{\gamma N \to VN}$ (i.e., the vector meson photoproduction amplitude
used in Eq.~(\ref{pia})) as
$ f_{\gamma N \to VN} = \frac{\sqrt{\pi \alpha_{em}}}{\gamma_V} f_{MN} $.
$\alpha_{em}$ is the fine structure constant.
$\gamma_V$
is the photon to vector meson coupling constant as described in VMD model
\cite{saku}. The values of $\gamma_\rho$ and $\gamma_\omega$, extracted
from the measured width of $V \to e^+e^-$ \cite{olive}, are 2.48 and 8.53
respectively \cite{sibir2}.
For
the pion nucleon scattering amplitude, the energy dependent experimentally
determined values of $f_{\pi^\pm N}$ are listed in Ref.~\cite{barn}. The
$\pi^0$ meson nucleon scattering amplitude $f_{\pi^0 N}$ is estimated
(using isospin algebra) as
$ f_{\pi^0 N} = \frac{1}{2} [ f_{\pi^+ N} + f_{\pi^- N} ] $. They are used
to evaluate the pion nucleus optical potential.

The calculated $ab$ (i.e., either $\pi^0\gamma$ or $\pi^+\pi^-$) invariant
mass distribution spectra are presented afterwards in the figures, 
where
the large-dashed curve represents the cross section due to $\omega$ meson.
The cross section because of the $\rho^0$ meson is shown by the short-dashed
curve.
The
dot-dashed curve arises because of the incoherent contribution of these
mesons to the cross section of the reaction, i.e., the summation of the
cross sections of the coherent $(\gamma, \rho^0 \to ab)$ and
$(\gamma, \omega \to ab)$ reactions on the nucleus.
The
solid curve illustrates the coherently added cross sections of these
reactions, i.e., the amplitudes of the reactions are added to get the
cross section. The dot-dot-dashed curve shows the contribution to the
cross section arising due to the $\rho-\omega$ interference.

\subsection { $(\gamma, V \to \pi^0\gamma)$ reaction }

The cross section for the $\pi^0\gamma$ invariant mass distribution 
in the coherent $(\gamma,V)$ reaction on the isoscalar nuclei are calculated
using the formalism (developed in the previous section)
where
the particles $a$ and $b$ in the final state are replaced by $\pi^0$ and
$\gamma$ bosons respectively. The $V\to \pi^0\gamma$ decay is described by
the Lagrangian \cite{lyka, teje}:
\begin{equation}
{\cal L}_{V\pi\gamma}= \frac{f_{V\pi\gamma}}{m_\pi}
\varepsilon_{\alpha \beta \delta \sigma}
\partial^\alpha A^\beta {\bf \pi} \cdot \partial^\delta {\bf V}^\sigma,
\label{lgnV}
\end{equation}
where $f_{V\pi\gamma}$ denotes the $V\pi\gamma$ coupling constant.
The
width of this decay \cite{das4} is given by
\begin{equation}
\Gamma_{V \to \pi^0\gamma} (m) = \Gamma_{V \to \pi^0\gamma} (m_V)
\left [ \frac{ {\tilde k} (m) }{ {\tilde k} (m_V) } \right ]^3
\Theta (m - m_\pi),
\label{wdpg}
\end{equation}
where ${\tilde k} (m)$ is the momentum of the pion originating due to
the vector meson of mass $m$ decaying at rest.
Since
the final state interaction of the $\gamma$ boson is negligible,
$F_{\gamma, V\to ab}$ in Eq.~(\ref{fve0}) can be expressed as
\begin{equation}
F_{\gamma, V\to \pi^0\gamma} = \int d{\bf r} \int^\infty_z dz^\prime
D^{(-)*}_{ {\bf k}_{\pi^0} } ( {\bf b}, z^\prime )
D_{{\bf k}_V} (m; {\bf b}, z^\prime, z)
e^{i({\bf k}_\gamma - {\bf k}_V).{\bf r}} \Pi_{\gamma A \to VA} ({\bf r}).
\label{fve2}
\end{equation}

The double differential cross section for the correlated $\pi^0\gamma$
invariant mass $m$ distribution in the coherent
$(\gamma,V \to \pi^0\gamma)$ reaction on a nucleus, using Eq.~(\ref{dsc}),
can be written as
\begin{equation}
\frac{d\sigma}{dmd\Omega_V} = P_{\pi^0\gamma}
| \sum_{V=\rho^0, \omega} \Gamma^{1/2}_{V \to \pi^0\gamma} (m)
F_{\gamma, V\to \pi^0\gamma } |^2,
\label{dscn}
\end{equation}
where $P_{\pi^0\gamma}$ in this equation arises because of the phase-space
of the reaction:
$P_{\pi^0\gamma} = \frac{ 1 }{ (2\pi)^3 }
 \frac{ k^2_Vm^2E_{A^\prime} }{ E_\gamma|k_VE_i-k_\gamma cos\theta_V E_V| }$.
$E_{A^\prime}$ denotes the energy of the recoiling nucleus.
The
$\pi^0\gamma$ invariant mass distribution spectra are calculated for the
fixed $\pi^0$ meson emission angle ($\theta_{\pi^0}=1^0$) in the
multi-GeV region.
The
bosons in the final state (i.e., $\pi^0$ and $\gamma$) arise due to the
decay of the forward ($\theta_V=0^0$) going vector meson (i.e., $\rho^0$
or $\omega$ meson) which is photoproduced coherently in the isoscalar
nucleus.

The calculated $\pi^0 \gamma$ invariant mass distribution spectra for
$^{12}$C nucleus (a isoscalar nucleus) are shown in Fig.~\ref{Fg1}. The
beam energy $E_\gamma$ is taken equal to 3 GeV. The density distribution
$\varrho ({\bf r})$ of this nucleus is described by the Harmonic Oscillator
Gaussian form \cite{andt}, i.e.,
\begin{equation}
\varrho ({\bf r})= \varrho_0 [1+w(r/c)^2] e^{-(r/c)^2};
\label{dnC}
\end{equation}
with $w=1.247$, $c=1.649$ fm.
Fig.~\ref{Fg1}
shows that the cross section at the peak due to $\omega$ meson
(large-dashed curve) is distinctly largest. It appears at the pole mass of
$\omega$ meson, i.e., $m\sim 0.78$ GeV. In this region, the cross section
because of the $\rho^0$ meson (short-dashed curve) is negligibly small
compared to the previous.
The
peak cross section, as shown in Fig.~\ref{Fg1}(a), is increased by
$\approx 12\%$ because of the inclusion of the $\rho^0$ meson
contribution in the calculated cross section.

It should be mentioned that the distinct $\rho^0 - \omega$ interference
in the $\pi^+\pi^-$ and $e^+e^-$ decay channels has been seen beyond the
peak of the cross section \cite{biggs, behr}. To explore that in the
$\pi^0\gamma$ decay channel, the calculated cross section away from the
peak region is presented in Fig.~\ref{Fg1}(b) for $^{12}$C nucleus.
The
$\rho -\omega$ interference is distinctly visible in this figure in the
regions of $m$ below ($m<0.76$ GeV) and beyond ($m>0.80$ GeV) its value at
the peak position.
This
figure also shows that the enhancement in the calculated cross section due
to the inclusion of $\rho$ meson in it (as shown in Fig.~\ref{Fg1}) is much
larger than that occurs at the peak of the cross section, and the quoted
enhancement increases as $m$ is away from its value at the peak position.
Fig.~\ref{Fg1}(b)
shows that the contribution of $\rho$ meson to the cross section is
comparable to that of the $\omega$ meson at $m$(GeV) $\sim$ 0.64 and 0.92.
In
these regions of $m$, the calculated cross section is increased by
$\sim 100\%$ due to the $\rho -\omega$ interference.
In
fact, the cross section is increased drastically (a factor of $\sim 3.5$)
because of the $\rho$ meson, resulting a remarkable change in
the shape of $\pi^0\gamma$ invariant mass distribution spectrum in the
regions few hundred MeV away from its peak.

To look for the contribution of the $\rho^0$ meson at higher energy, the
$\pi^0\gamma$ invariant mass distribution spectra of the considered
reaction are calculated at $E_\gamma =5$ GeV.
The
calculated results, presented in Fig.~\ref{Fg2}, illustrates that the
$\pi^0\gamma$ invariant mass $m$ distribution spectra at 5 GeV are
qualitatively similar to those at 3 GeV, see Fig.~\ref{Fg1}(b).
The
quoted interference and the change in the spectral shape (because of the
$\rho$ meson) in the region of $m$ few hundred MeV away from its value
at the peak position is also noticeable in this figure.
The
magnitude of the cross section is significantly increased with the beam
energy. At the peak, it is enhanced by a factor of 3 due to the increase in
the beam energy from 3 to 5 GeV.
The
cross section is increased from 47.3 $\mu$b/(GeV sr) to 0.12 mb/(GeV sr) at
$m=0.64$ GeV, and it is increased from 59.52 $\mu$b/(GeV sr) to 0.29
mb/(GeV sr) at $m=0.92$ GeV because of the quoted enhancement in the beam
energy.

Since large cross section away from the $\omega$-meson peak region in the
$\pi^0\gamma$ invariant mass distribution spectrum is preferable to study
the $\rho -\omega$ interference, the heavy nucleus should be considered for
it.
Therefore,
the cross section of the quoted reaction for $^{40}$Ca (a heavier isoscalar
nucleus) is calculated at 5 GeV. The density distribution
$\varrho ({\bf r})$ of this nucleus can be described by 3-parameter Fermi
distribution function \cite{andt}:
\begin{equation}
\varrho ({\bf r})
= \varrho_0 \frac{ 1+w(r/c)^2 }{ 1+exp[(r-c)/z] };
\label{dnCa}
\end{equation}
with $w=-0.1017$, $c=3.6685$ fm and $z=0.5839$ fm \cite{andt}.
The
calculated spectra, as shown in Fig.~\ref{Fg3}, are qualitatively similar
to those presented in the previous figures. The cross section for $^{40}$Ca
is significantly large compared to that for $^{12}$C nucleus. As shown in
this figure, the cross section at $m=0.64$ GeV is 1.15 mb/(GeV sr) and
it is 1.91 mb/(GeV sr) at $m=0.92$ GeV.

The ratio of the amplitude of $\rho^0$ meson to that of the $\omega$ meson
is given by
$
\xi = \frac
{ \Gamma^{1/2}_{\rho \to \pi^0\gamma} | F_{\gamma, \rho \to \pi^0\gamma}| }
{ \Gamma^{1/2}_{\omega \to \pi^0\gamma} | F_{\gamma, \omega \to \pi^0\gamma}| }.
$,
see Eq.~(\ref{dscn}). The value of $\xi$ for $^{12}$C nucleus (calculated
at $m_{\rho^0}$) is $\simeq 1.17 \times 10^{-1}$ and the relative phase
$\phi$ of these mesons is equal to $61^0$ at $E_\gamma =3$ GeV.
At
higher energy, i.e., $E_\gamma =5$ GeV, the calculated values of $\xi$ and
$\phi$ for this nucleus are $1.24 \times 10^{-1}$ and $63^0$ where as those
for $^{40}$Ca nucleus are found equal to $1.28 \times 10^{-1}$ and
$\phi \approx 61^0$.

\subsection { $(\gamma, V \to \pi^+\pi^-)$ reaction }

The formalism already described is used to calculate the cross section
for the $\pi^+\pi^-$ invariant mass distribution spectra in the coherent
$(\gamma,V)$ reaction on the nucleus.
In
this reaction, the vector meson $V$ decay products $a$ and $b$ (appearing
in the formalism) are $\pi^+$ and $\pi^-$ respectively. The Lagrangian,
which describes $V\to \pi^+\pi^-$ decay \cite{saku, eric}, is given by
\begin{equation}
{\cal L}_{V\pi^+\pi^-}=
g_{V\pi\pi} {\bf V}^\mu \cdot ( {\pi} \times \partial_\mu {\bf \pi} ),
\label{lppV}
\end{equation}
where $g_{V\pi\pi}$ denotes the $V\pi\pi$ coupling constant. The
parameterized form of this decay width \cite{das4} can be written as
\begin{equation}
\Gamma_{V \to \pi^+\pi^-} (m) = \Gamma_{V \to \pi^+\pi^-} (m_V)
\left ( \frac{m_V}{m} \right )
\left [ \frac{ {\tilde k} (m) }{ {\tilde k} (m_V) } \right ]^3
\Theta (m - 2m_\pi),
\label{wdpp}
\end{equation}
${\tilde k} (m)$ is defined in Eq.~(\ref{wdpg}).
The
form of $F_{\gamma, V\to ab}$ in Eq.~(\ref{fve0}) for $V \to \pi^+\pi^-$
decay is given by
\begin{equation}
F_{\gamma, V\to \pi^+\pi^-} = \int d{\bf r} \int^\infty_z dz^\prime
D^{(-)*}_{ {\bf k}_{\pi^+} } ( {\bf b}, z^\prime )
D^{(-)*}_{ {\bf k}_{\pi^-} } ( {\bf b}, z^\prime )
D_{{\bf k}_V} (m; {\bf b}, z^\prime, z)
e^{i({\bf k}_\gamma - {\bf k}_V).{\bf r}} \Pi_{\gamma A \to VA} ({\bf r}).
\label{fve4}
\end{equation}

\begin{figure}[h]
%\begin{figure}
\begin{center}
\centerline {\vbox {
%\psdraft
\psfig{figure=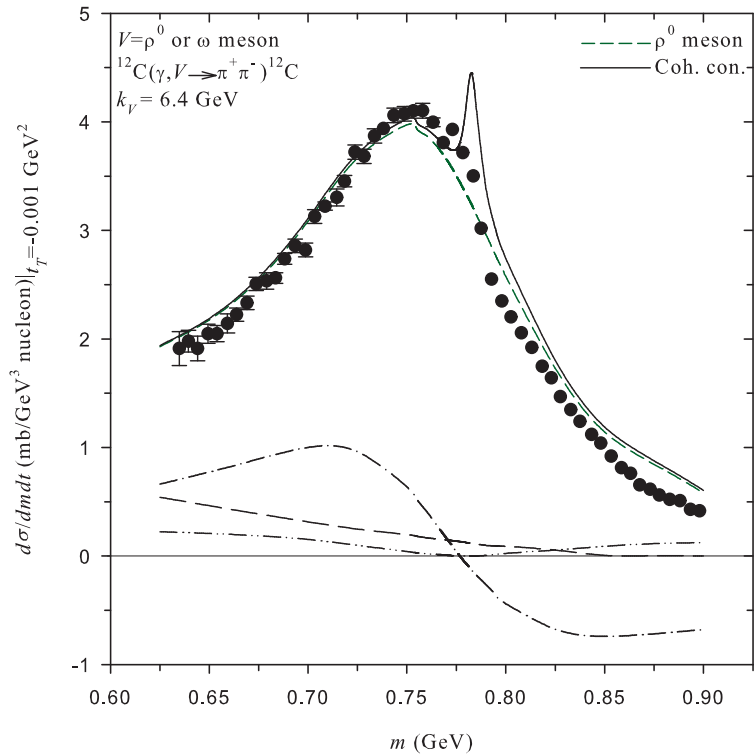,height=10.0 cm,width=10.0 cm}
}}
\caption{
(color online). The calculated $\pi^+\pi^-$ invariant mass $m$ distribution
spectra in the coherent $(\gamma,V)$ reaction on $^{12}$C nucleus.
The
data along with the background curves are taken from Ref.~\cite{alven2}. As
mentioned in it, the background curves are due to the interference with
non-resonant $\pi\pi$ emission (dot-dashed curve), the non-resonant $\pi\pi$
emission (dot-dot-dashed curve), and other background (large-dashed curve).
The
calculated results (added with the backgrounds) are compared with the data.
The solid curve arises because of the contribution of the $\omega$ meson
coherently added to that of the $\rho^0$ meson (short-dashed curve).
}
\label{Fg4}
\end{center}
\end{figure}

The double differential cross section $\frac{d\sigma}{dmdt}$ for the
$\pi^+\pi^-$ invariant mass $m$ distribution of the considered reaction
on $^{12}$C nucleus is calculated to compare that with the data for the
vector meson momentum $k_V=6.4$ GeV and the transverse four-momentum
transfer $t_T=-0.001$ GeV$^2$.
The
expression for $\frac{d\sigma}{dmdt}$, using Eq.~(\ref{dsc}), can be
written as
\begin{equation}
\frac{d\sigma}{dmdt} = P_{\pi^+\pi^-}
| \sum_{V=\rho^0, \omega} m^{1/2}_V \Gamma^{1/2}_{V \to \pi^+\pi^-} (m)
F_{\gamma, V\to \pi^+\pi^- } |^2,
\label{dscc}
\end{equation}
with
$P_{\pi^+\pi^-} = \frac{ \pi }{ (2\pi)^3 }
\frac{ k_VmE_{A^\prime} }{ E^2_\gamma|k_VE_i-k_\gamma cos\theta_V E_V| }$.
The
calculated ratio $\xi$ of the amplitudes of $\rho^0$ and $\omega$ mesons
and the relative phase $\phi$ of them are found equal to
$8.66 \times 10^{-2}$ and $\approx 65^0$ respectively.
The
measured $\pi^+\pi^-$ invariant mass $m$ distribution spectrum due to the
decay of vector meson and backgrounds, taken from Ref.~\cite{alven2}, are
presented in Fig.~\ref{Fg4}. The calculated spectra (added with the
backgrounds) are also shown in this figure.
The
short-dashed curve represents the cross section due to $\rho^0$ meson. The
solid curve arises because of the contribution of the $\omega$ meson
coherently added to the previous.
This
figure shows that the calculated spectrum due to $\rho$ meson is well
accord with the data. The change in this spectrum due to the inclusion
(coherently) of the $\omega$ meson contribution is insignificant except
a sharp peak appears at the pole mass of this meson, i.e.,
$m_\omega = 782.65$ MeV.

\begin{figure}[h]
%\begin{figure}
\begin{center}
\centerline {\vbox {
%\psdraft
\psfig{figure=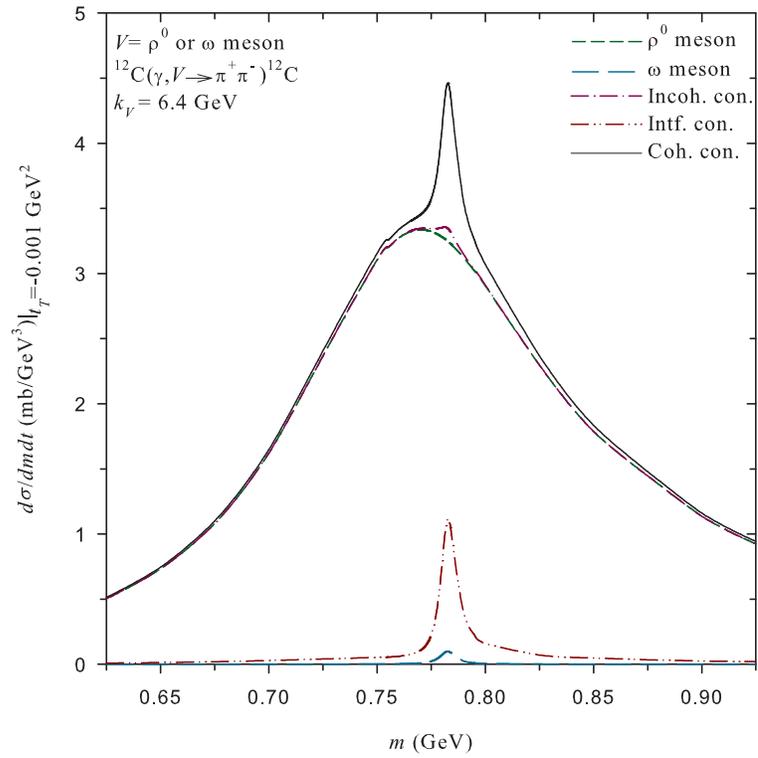,height=10.0 cm,width=10.0 cm}
}}
\caption{
(color online). The calculated $\pi^+\pi^-$ invariant mass $m$ distribution
spectra originating due to the decay of $\rho^0$ and $\omega$ mesons
photo-produced coherently in $^{12}$C nucleus. Various curves appear in it
are illustrated in the text.
}
\label{Fg5}
\end{center}
\end{figure}

It should be mentioned that the decay width
$\Gamma_{\omega \to \pi^+\pi^-} (m_\omega)$, equal to 0.13 MeV, is
negligibly small compared to $\Gamma_{\rho^0 \to \pi^+\pi^-} (m_{\rho^0})$,
i.e., $\sim$ 149 MeV.
Therefore,
a peak due to $\omega$ meson is not expected in the $\pi^+\pi^-$ invariant
mass distribution spectrum. To explore the origin of the sharp peak
appearing at the $\omega$ meson pole mass (i.e., $m=m_\omega$) in
Fig.~\ref{Fg4},
the
cross sections due to $\rho^0$ and $\omega$ mesons (along with their
interference) are plotted in Fig.~\ref{Fg5}.
This
figure shows that the cross section due to $\omega$ meson (large-dashed
curve) is negligibly small compared to that because of the $\rho^0$ meson
(short-dashed curve)
but
the cross section due to the $\rho-\omega$ interference
(dot-dot-dashed curve) is very much significant at $m = m_\omega$ in the
$\pi^+\pi^-$ invariant mass distribution spectrum.
Therefore,
the sharp peak distinctly visible at the $\omega$ meson pole mass in
Figs.~\ref{Fg4} and \ref{Fg5} arises because of the quoted interference.

\section{Conclusions}

The $\rho -\omega$ interference is studied in the $\pi^0\gamma$ and
$\pi^+\pi^-$ decay channels of the $\rho^0$ and $\omega$ mesons which are
considered to produce coherently in the photo-induced reaction on the
isoscalar nuclei.
The
distinctly dominant contribution to the cross section of the $\pi^0\gamma$
invariant mass distribution arises due to the $\omega \to \pi^0\gamma$ in
the peak region.
Therefore,
the $\rho^0 \to \pi^0\gamma$ channel can be ignored for studying the
physical phenomena associated with the $\omega$ meson (e.g., the
modification of this meson in a nucleus) in the peak region.
Few
hundred MeV away from this region, the $\rho^0 \to \pi^0\gamma$ channel is
very important because the contribution of this channel to the cross section
is comparable to that of $\omega \to \pi^0\gamma$ channel.
The
cross section is also significantly increased in these regions due to the
$\rho -\omega$ interference.
Therefore,
the cross section of the $\pi^0\gamma$ invariant mass distribution spectrum
is drastically increased and the shape of it is remarkably changed due to
$\rho$ meson in the above mentioned regions of the spectrum.
The
cross section increases with the size of the nucleus and beam energy.
The
measureable cross section exits (specifically for the heavy nucleus and
high energy) in the regions of $\pi^0\gamma$ invariant mass distribution
spectrum where the contribution of the $\rho$ meson and the $\rho-\omega$
interference is significant.

The spectrum calculated for the $\pi^+\pi^-$ invariant mass distribution
due to $\rho^0$ meson reproduces the data reasonably well except in the
$\omega$ meson peak region.
The
calculated results show that the cross section because of the $\omega$
meson is insignificant compared to that due to the $\rho$ meson.
The
cross section due to the $\rho^0-\omega$ interference is significantly
large at the pole mass of $\omega$ meson, resulting a sharp peak at
this mass appearing in the $\pi^+\pi^-$ invariant mass distribution
spectrum.

\section{Acknowledgement}

The author takes the opportunity to thank the anonymous referee for
giving valuable comments which helped to improve the quality of the
paper.
Drs. A. Saxena and B. K. Nayak are gratefully acknowledged for their
encouragement to work on the intermediate energy nuclear physics.

%\newpage

\newpage

\begin{figure}[h]
%\begin{figure}
\begin{center}
\centerline {\vbox {
%\psdraft
\psfig{figure=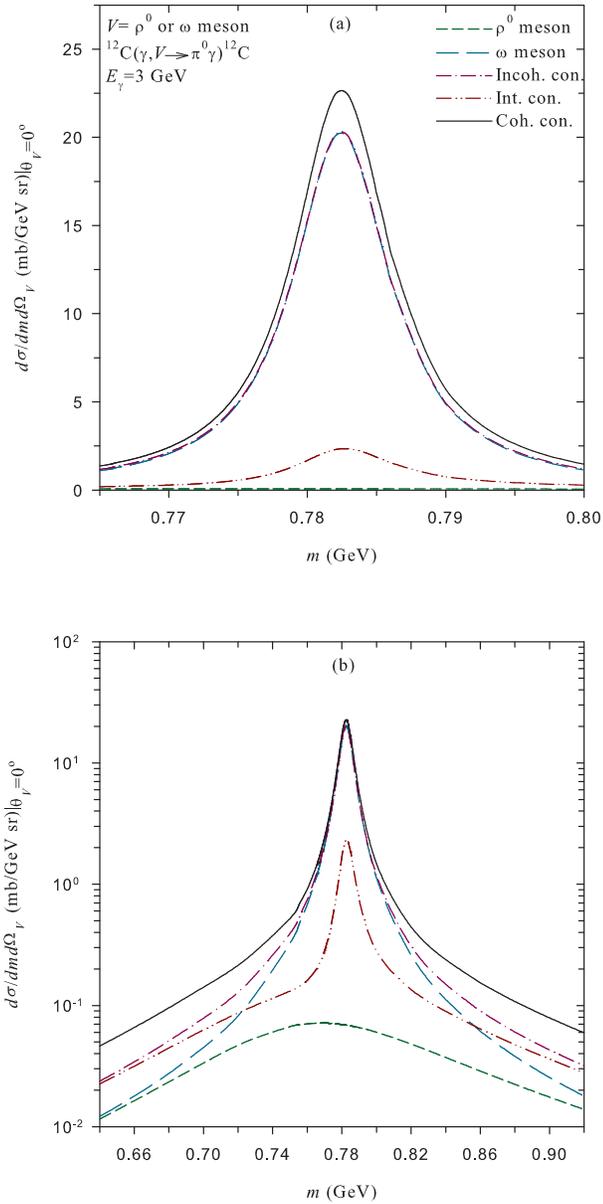,height=16.0 cm,width=8.0 cm}
}}
\caption{
(color online). The $\pi^0\gamma$ invariant mass $m$ distribution spectra
in the coherent $(\gamma,V)$ reaction on $^{12}$C nucleus. The enhancement
in the cross section due to $\rho^0$ meson is distinctly visible beyond the
peak region. See text for the explanation of various curves appearing in
the figure.
}
\label{Fg1}
\end{center}
\end{figure}

\begin{figure}[h]
%\begin{figure}
\begin{center}
\centerline {\vbox {
%\psdraft
\psfig{figure=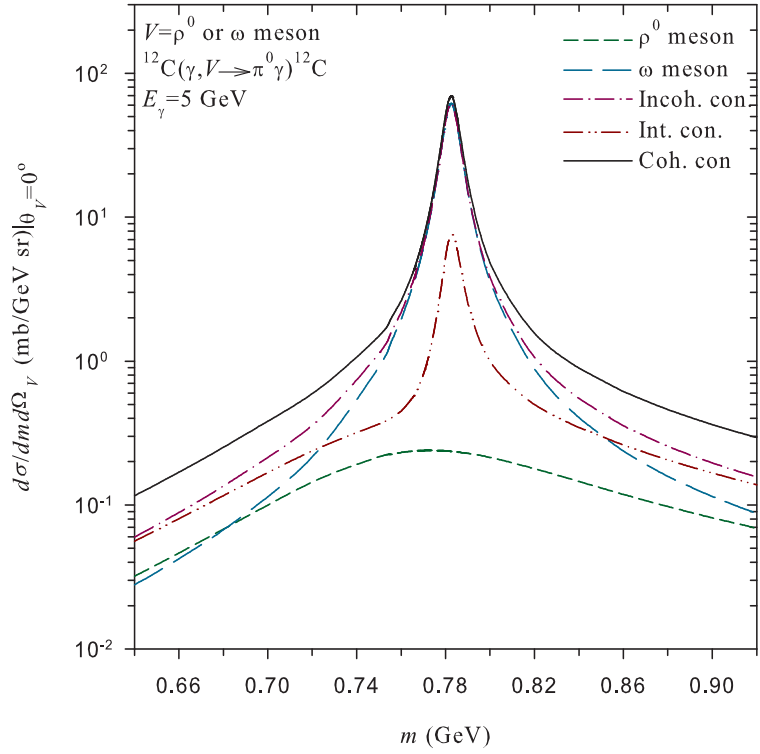,height=10.0 cm,width=10.0 cm}
}}
\caption{
(color online). Same as those presented in Fig.~\ref{Fg1}(b) but for the
beam energy taken equal to 5 GeV.
}                             
\label{Fg2}
\end{center}
\end{figure}

\begin{figure}[h]
%\begin{figure}
\begin{center}
\centerline {\vbox {
%\psdraft
\psfig{figure=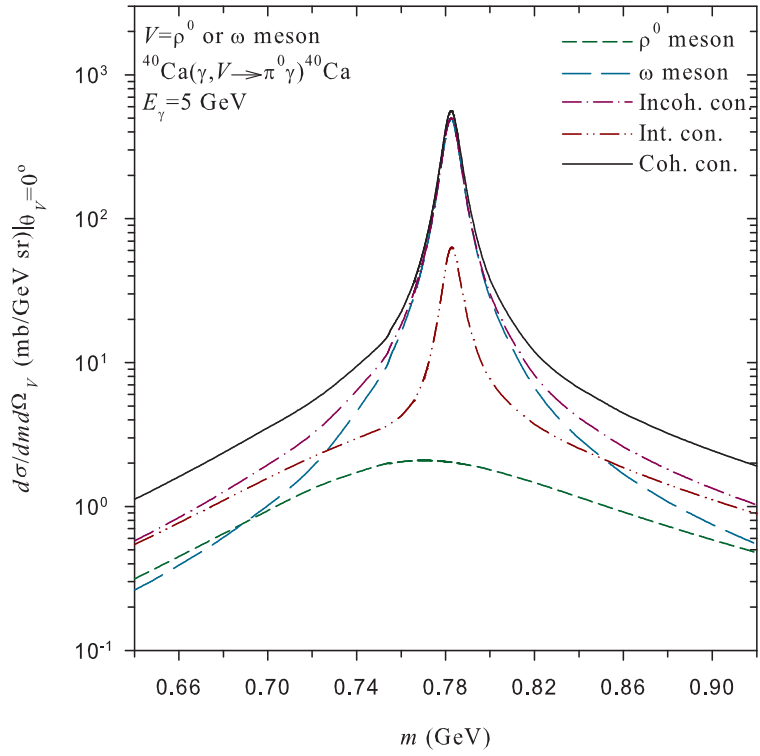,height=10.0 cm,width=10.0 cm}
}}
\caption{
(color online). Same as those presented in Fig.~\ref{Fg2} but for
$^{40}$Ca nucleus.
}
\label{Fg3}
\end{center}
\end{figure}

{\bf Figure Captions}
\begin{enumerate}

\item
(color online). The $\pi^0\gamma$ invariant mass $m$ distribution spectra
in the coherent $(\gamma,V)$ reaction on $^{12}$C nucleus. The enhancement
in the cross section due to $\rho^0$ meson is distinctly visible beyond the
peak region. See text for the explanation of various curves appearing in
the figure.

\item
(color online). Same as those presented in Fig.~\ref{Fg1}(b) but for the
beam energy taken equal to 5 GeV.

\item
(color online). Same as those presented in Fig.~\ref{Fg2} but for
$^{40}$Ca nucleus.

\item
(color online). The calculated $\pi^+\pi^-$ invariant mass $m$ distribution
spectra in the coherent $(\gamma,V)$ reaction on $^{12}$C nucleus.
The
data along with the background curves are taken from Ref.~\cite{alven2}. As
mentioned in it, the background curves are due to the interference with
non-resonant $\pi\pi$ emission (dot-dashed curve), the non-resonant $\pi\pi$
emission (dot-dot-dashed curve), and other background (large-dashed curve).
The
calculated results (added with the backgrounds) are compared with the data.
The solid curve arises because of the contribution of the $\omega$ meson
coherently added to that of the $\rho^0$ meson (short-dashed curve).

\item
(color online). The calculated $\pi^+\pi^-$ invariant mass $m$ distribution
spectra originating due to the decay of $\rho^0$ and $\omega$ mesons
photo-produced coherently in $^{12}$C nucleus. Various curves appear in it
are illustrated in the text.

\end{enumerate}

\newpage
%\vspace{1 cm}

\newpage
%\vspace{1 cm}

\newpage
%\vspace{1 cm}

\newpage
%\vspace{1 cm}

\newpage
%\vspace{1 cm}

\end{document}